\documentclass[reprint,amsmath,superscriptaddress,showpacs,showkeys,amssymb,aps,apl]{revtex4-1}
\usepackage{graphicx,epsfig,multirow,dcolumn,bm}

\begin{document}
\title{Unusual Strain glassy phase in Fe doped Ni$_2$Mn$_{1.5}$In$_{0.5}$}

\author{R. Nevgi}
\author{K. R. Priolkar}
\affiliation{Department of Physics, Goa University, Taleigao Plateau, Goa 403206 India}

\begin{abstract}
Fe doped Ni$_2$Mn$_{1.5}$In$_{0.5}$, particularly, Ni$_2$Mn$_{1.4}$Fe$_{0.1}$In$_{0.5}$, despite having an incommensurate, modulated 7M martensitic structure at room temperature exhibits frequency dependent behavior of storage modulus and loss that obeys Vogel-Fulcher law as well as shows ergodicity breaking between zero field cooled and field cooled strain measurements just above the transition temperature. Both, frequency dependence and ergodicity breaking are characteristics of a strain glassy phase and occur due to presence of strain domains which are large enough to present signatures of long range martensitic order in diffraction but are non interacting with other strain domains due to presence of Fe impurity.
\end{abstract}

\maketitle

The term Glass describes a frozen state of a certain local order of a statistically disordered system in general. Glassy phase comprises of a wide range of systems which include amorphous, ferroic system, polymers, biological systems etc. \cite{Angell1995205702,Sokolov1996273}. A strain glass is a ferroelastic state with a short range ordering of elastic strain vector. It is analogous to spin glass and relaxor in ferromagnetic and ferroelectric states respectively \cite{Ren2012}. The evidence for the existence of strain glass phase in off stoichiometric NiTi martensitic binary alloy (Ni$_{50+x}$Ti$_{50-x}$, $x \geq$ 0.15) is first reported by \cite{Sarkar200595}. It is believed to be a result of sufficient doping of point defects (Ni at Ti site) suppressing the long range ordering of elastic strain vector (martensitic transformation) preferring a B2 ${\rightarrow}$ R (trigonal) over B2 ${\rightarrow}$ B19 (monoclinic) transformation. The strain glass is unable to reach the long range strain order and is locked in a state in which short range persists \cite{Sarkar200595,Zhang201184}.

Apart from Ni rich NiTi alloys, strain glassy phase has been found in other impurity doped alloys including Ti$_{50}$Ni$_{50-x}$D$_x$ (D = Fe, Co, Cr, Mn) \cite{Wang201058,Zhou201058} with D acting as an impurity. Similar situation is seen in Ti$_{50}$Pd$_{50-x}$Cr$_x$ \cite{Zhou200995} wherein a crossover from martensitic transition to stain glass transition is observed at critical doping concentration. The magnetic shape memory alloy Ni$_{55-x}$Co$_x$Fe$_{18}$Ga$_{27}$ also exhibits characteristics of strain glass transition at a critical Co level of 10${\%}$ \cite{Wang201298}.

In literature, a strain glassy phase has been clearly distinguished from pre-martensitic tweed formation \cite{Ren2012}. Strain glass exhibits a frequency dependent anomaly in its dynamical mechanical properties around the transition temperature $T_g$. The anomaly is the existence of a dip in the ac storage modulus curve and corresponding peak in loss ($\tan\delta$) curve. A frequency dependence of $T_g$ obeying the Vogel-Fulcher law is also seen \cite{Ren201090}. The other trait of glass transition which distinguishes it from pre-martensitic tweed is the existence of ergodicity breaking evidenced in zero field cooled (ZFC) and field cooled (FC) experiments. Strain glass transition is also characterized by an invariant crystal structure across the glass transition. The strain glassy phase is identified with formation of nano sized domains with frozen elastic strain vector, the long range structural order and consequently the crystal structure does not change.

Fe doping in martensitic Ni-Mn-In alloys results in suppression of $T_M$ and strengthening of ferromagnetic interactions \cite{Sharma2012111,Lobo2014116}. The suppression is rather rapid and is explained to be due to destruction of Mn - Ni - Mn antiferromagnetic interactions and formation of Fe - Fe ferromagnetic interactions due to site occupancy disorder \cite{Lobo2014116}. The question then arises is to whether Fe doping in martensitic Ni$_2$MnIn alloys also results in impeding long range ordering of elastic strain vector and formation of strain glass phase similar to the one observed in impurity doped NiTi alloys. We attempt to answer this question by studying the structure, thermal and frequency dependent elastic properties of Ni$_2$Mn$_{1.5-x}$Fe$_x$In$_{0.5}$ alloys. Here the undoped, Ni$_2$Mn$_{1.5}$In$_{0.5}$ is martensitic below 422K and our results show that Fe doping not only results in decrease in $T_M$ but also in formation of an unusual strain glassy phase.

The synthesis of Ni$_2$Mn$_{1.5-x}$Fe$_x$In$_{0.5}$ ($x$ = 0, 0.025, 0.05, 0.075, 0.1, 0.15, 0.2 ) was carried out by arc melting in argon atmosphere by taking stoichiometric proportions of each constituent elements. The beads of each alloy formed were melted several times by flipping over to ensure homogeneity. A part of each bead was cut into suitable sizes and the remaining powdered. The powder covered in tantalum foil and  the pieces were encapsulated in evacuated quartz tube, annealed at 750$^\circ$C for 48 hours and subsequently quenched in ice cold water. Room temperature x-ray diffraction patterns of the powdered alloys were recorded using Mo K$_\alpha$ radiation in the angular range of 10$^\circ$ to 70$^\circ$ to obtain structural information. The prepared compositions were checked by SEM-EDX measurements. All alloys were found to have compositions within 2 to 5\% of stoichiometric values. To confirm martensitic transformation temperatures, differential scanning calorimetry (DSC) and four probe resistivity measurements were performed. DSC measurements were accomplished using Shimadzu DSC-60 on 6 to 7 mg pieces of each alloy crimped in aluminium pans and resistivity measurements were concluded using Oxford Instruments Optistat DNV on rectangular pieces of about 9.7 mm in length. Frequency dependent measurements of AC storage modulus and internal friction ($\tan \delta$) were carried out using a dynamic mechanical analyzer (Q800, TA Instruments). Measurements were carried out as a function of temperature, using 3 point bending mode by applying a small AC stress that generated a maximum displacement of 5$\mu$m at different frequencies in the range 0.1Hz to 7Hz on samples cut in a rectangular bars of (10mm x 3mm x 1mm) dimensions.

\begin{figure}[h]
\begin{center}
\includegraphics[width=\columnwidth]{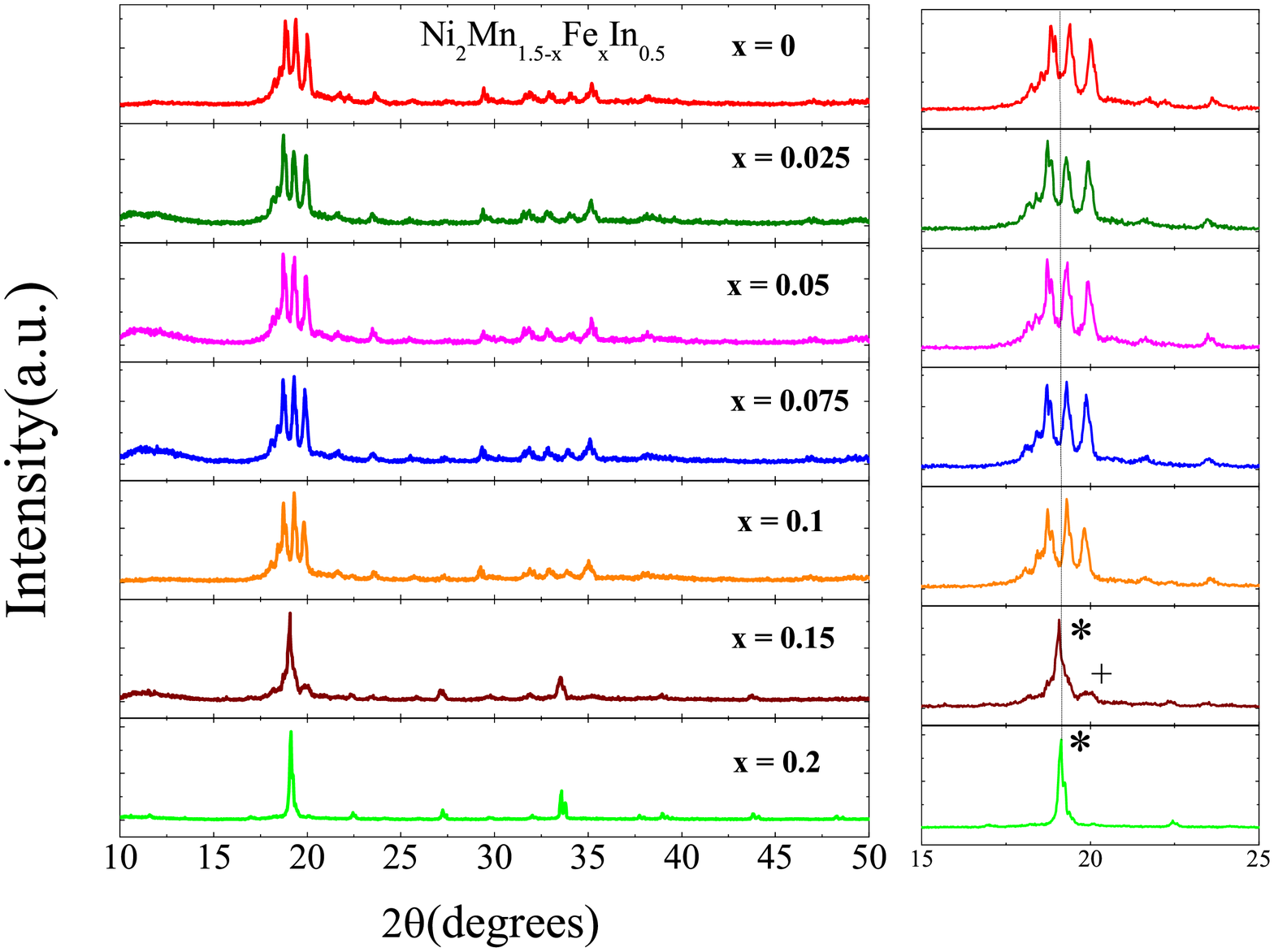}
\caption{X-ray diffraction patterns of Ni$_2$Mn$_{1.5-x}$Fe$_x$In$_{0.5}$ indicating incommensurate 7M modulated martensitic structure in the alloys $x$ = 0, 0.025, 0.05, 0.075, 0.1. The phase co-existence can be clearly seen in the composition x = 0.15 in the form of cubic and martensitic peaks marked as $*$ and $+$ respectively while x = 0.2 shows an almost grown cubic phase.}
\label{fig:XRD}
\end{center}
\end{figure}

\begin{figure}[h]
\begin{center}
\includegraphics[width=\columnwidth]{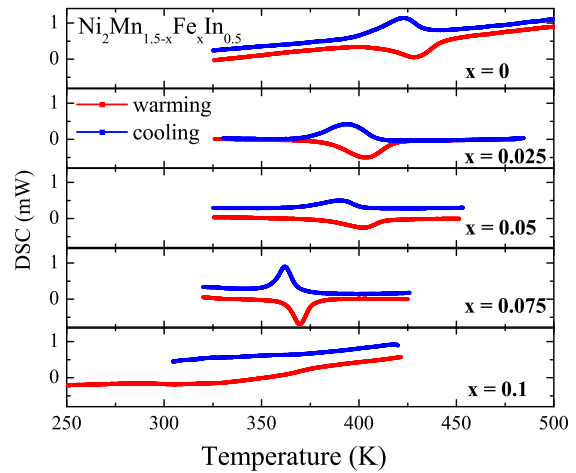}
\caption{Differential scanning calorimetry plots during warming and cooling cycles in Ni$_2$Mn$_{1.5-x}$Fe$_x$In$_{0.5}$ ($x$ = 0, 0.025, 0.05, 0.075, 0.1).}
\label{fig:DSC}
\end{center}
\end{figure}

The room temperature x-ray diffraction patterns presented in Fig.\ref{fig:XRD} show a modulated martensitic structure for samples with $x \le  0.1$ indicating that these alloys undergo martensitic transformation at a temperature $T_M >$ room temperature. On the other hand, compositions $x$ = 0.15 and $x$ = 0.2 show a two phase pattern consisting of the cubic austenite and modulated martensite phases (indicate in Fig. \ref{fig:XRD} with $*$ and $+$ signs respectively). Estimated phase fractions of cubic and martensitic phases from Lebail refinement were obtained respectively as 57.8:47.2 for $x$ = 0.15 and 89.5:10.5 for $x$ = 0.2 alloys. This indicates that Fe doping results in the growth of cubic phase at the expense of martensitic phase. The rate of growth of cubic phase suggests a possibility of existence of a minor cubic phase even in $x$ = 0.1 alloy but present diffraction measurements did not detect presence of any impurity phase. The austenite  to martensite transition temperature was determined through the DSC measurements which are depicted in Fig.\ref{fig:DSC}. The transformation to martensitic state reflects as an exothermic and endothermic peak during warming and cooling cycles respectively. The hysteresis in positions of the peaks during warming and cooling confirms the first order nature of the transformation.

The sensitivity of $T_M$ to Fe content is evident through the fact that the transition temperature decreases sharply with small doping concentration of Fe  at the expense of Mn in Ni$_2$Mn$_{1.5}$In$_{0.5}$. While the undoped alloy transforms from austenitic to martensitic state at 422K, in Ni$_2$Mn$_{1.425}$Fe$_{0.075}$In$_{0.5}$ the transformation occurs at 355K. Interestingly, the composition $x$ = 0.1 seems to show a broad feature over a otherwise sharp transition observed in the DSC when a material transforms via a first order transition (Fig.\ref{fig:DSC}). Such a nearly vanishing DSC peak has been attributed to short range ordering of strain vector across the ferroelastic transition \cite{Ren2012}. It may be noted the compositions with $x$ = 0.15 and $x$ = 0.2 do not show any transition down to lowest measured temperature despite having sizable ($> 10\%$) fraction of martensitic phase. Recently, crystallization of  strain glass via a isothermal transformation has been reported in Ni rich NiTi alloys \cite{PhysRevLett.114.055701}. In order to check such a possibility of isothermal growth of martensitic phase in these alloys, DSC measurements were performed on $x$ = 0.1 alloy using the same procedure as described in Ref. \onlinecite{PhysRevLett.114.055701}. No growth of a heat loss peak indicating appearance of martensitic phase was observed in these measurements. Therefore the observed weak feature in DSC of $x$ = 0.1 alloy could possibly be due a continuous transformation due compositional disorder or due to existence of more than one structural phases wherein one of them is martensitic and its transformation is inhibited by the other impurity phases. Compositional disorder can be ruled out in $x$ = 0.1 as the EDX measurements report its composition to be Ni$_{2.00}$Mn$_{1.36}$Fe$_{0.14}$In$_{0.50}$ which is quite close to the prepared composition.

\begin{figure}[h]
\begin{center}
\includegraphics[width=\columnwidth]{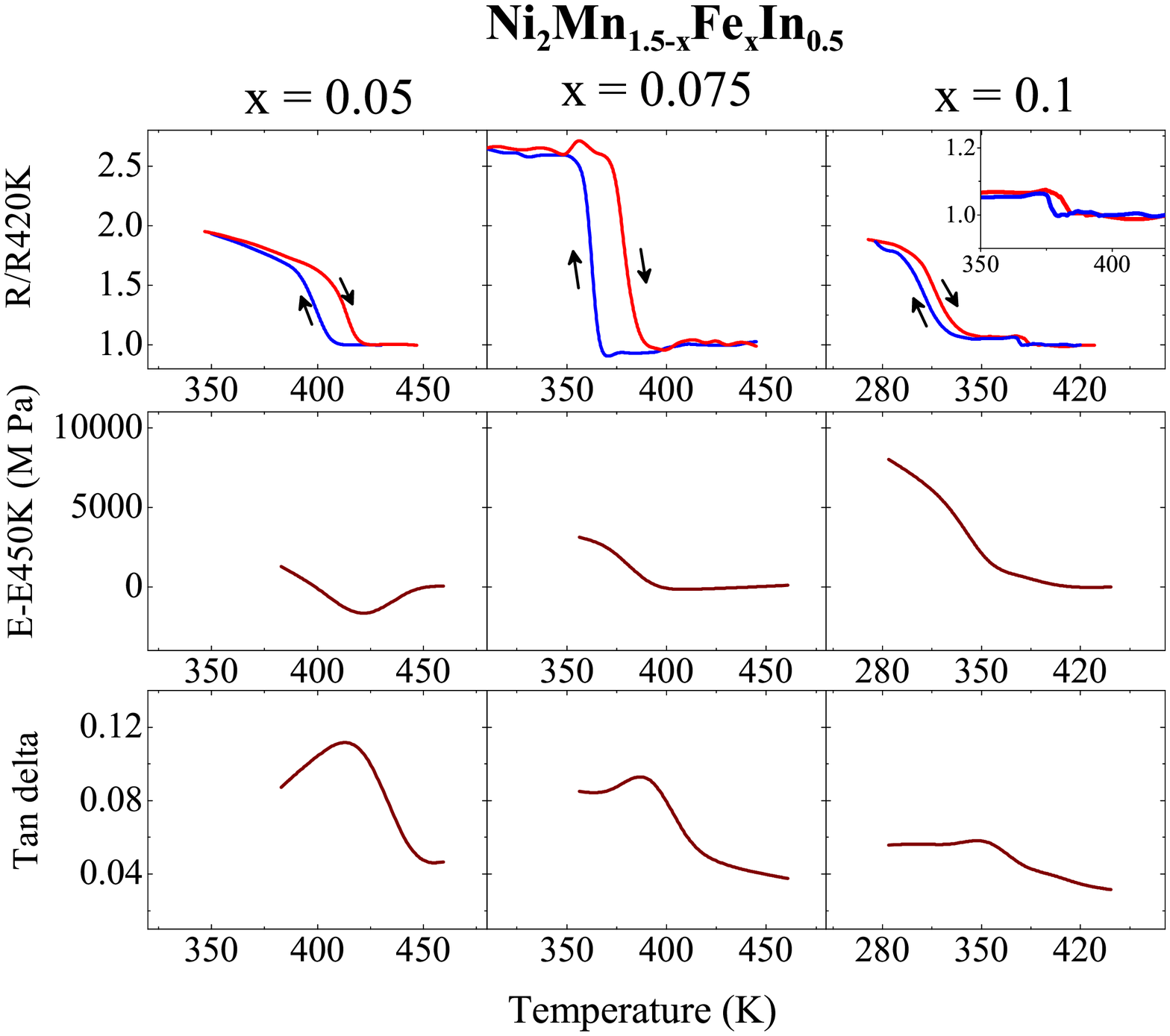}
\caption{The temperature dependence of the normalized resistance and the ac storage modulus and $\tan\delta$ in Ni$_2$Mn$_{1.5-x}$Fe$_x$In$_{0.5}$}
\label{fig:DMA+RES}
\end{center}
\end{figure}

Resistivity measurements  Fig.\ref{fig:DMA+RES} carried out on the compositions $x$ = 0.05 and $x$ = 0.075 expectedly show a sharp rise in resistance values as a signature of first order transition in the same temperature range as DSC measurements. In case of $x$ = 0.1 alloy, a much slower rise in resistivity is observed around 350K is observed consistent with the broad transition in DSC thermogram. Additionally a weak first order transition is seen at $\sim$ 380K in the resistivity measurements of $x$ = 0.1 alloy (see inset of Fig.\ref{fig:DMA+RES}(g) for clarity). While the weak first order transition at 380K could be due to martensitic transformation which explains the observation of modulated structure in XRD at room temperature, the broad transition at 350K could be due to short range order of elastic strain vector. Presence of two transitions, one hinting at long range martensitic order and another one pointing to some sort of glassy phase transition indicates presence of phase co-existence in this alloy and hence further on in this letter, we focus our attention on alloy compositions $x \le 0.1$.

Impurity doping in martensites are known to be initiators of a conjugate transition from austenitic phase to a strain glass phase which is a frozen disordered state of short range ordered strain vectors. The characteristics of such a glassy state are (a) frequency dependence of ac modulus/loss exhibiting behavior according to Vogel-Fulcher relation, (b) ergodic symmetry breaking between ZFC and FC curves around the glass transition temperature, (c) invariance of average structure and (d) existence of short range order in the glassy state. Here, Ni$_2$Mn$_{1.4}$Fe$_{0.1}$In$_{0.5}$ though exhibits a long range ordered martensitic structure at room temperature shows a vanishingly small DSC peak and a broad transition in resistivity at about 350K which are considered to be signatures of glassy dynamics. To understand this seemingly paradoxical situation better, AC storage modulus and internal friction or loss were measured as a function of temperature at several different frequencies between 10Hz to 0.1Hz and compared with other alloys with lesser Fe content. Behavior of storage modulus and loss ($\tan\delta$) for three alloy compositions, $x$ = 0.05, 0.075 and 0.1 at a characteristic frequency of 5Hz are presented in Fig. \ref{fig:DMA+RES}.

\begin{figure}[h]
\begin{center}
\includegraphics[width=\columnwidth]{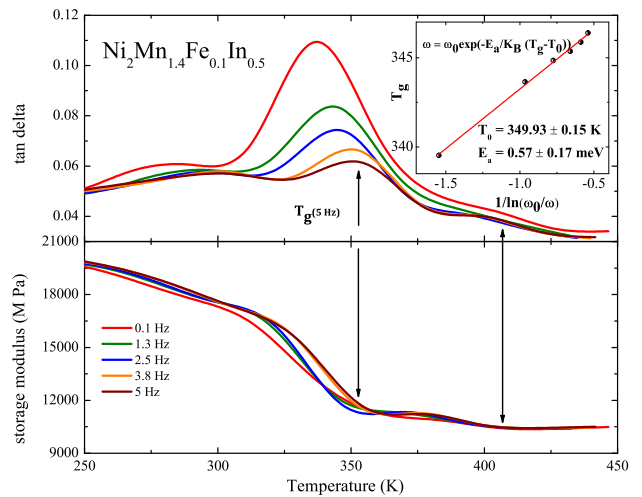}
\caption{The frequency dependent behavior of ac storage modulus and $\tan\delta$ observed in Ni$_2$Mn$_{1.4}$Fe$_{0.1}$In$_{0.5}$. Inset shows the logarithmic dependence of the peak in $\tan\delta$ along with a best fit to Vogel Fulcher relation (solid line).}
\label{fig:VF}
\end{center}
\end{figure}

Temperature evolution of ac storage modulus and $\tan\delta$ respectively exhibit a dip followed by a sharp increase and a peak at around $T_M$ in all alloy compositions up to $x$ = 0.075 (Fig.\ref{fig:DMA+RES}). The intensity of the peak as well as the sharpness of the rise however, decreases with increasing Fe content. In Ni$_2$Mn$_{1.4}$Fe$_{0.1}$In$_{0.5}$, the sharp anomaly converts to a broad feature followed by a slow rise of storage modulus at about 350K which is in good agreement with the results obtained from DSC and resistivity measurements. The peak in $\tan\delta$ observed at the same temperature and classified as $T_g$ exhibits a frequency dependence as can be clearly seen in Fig. \ref{fig:VF}. Such a frequency dependence is absent in all other alloys with lower Fe content. A plot of $T_g$ versus $\log$(frequency) presented in the inset of Fig.\ref{fig:VF} can be fitted to the Vogel Fulcher law, $\omega = \omega_0\exp[-E_a/k_B(T - T_0)]$, where $E_a$ is the activation energy and $T_0$ is the "ideal glass" temperature. This indicates a possibility of a glassy transition in Ni$_2$Mn$_{1.4}$Fe$_{0.1}$In$_{0.5}$. The relative shift of glass transition temperature is assessed by a parameter $k = \frac{\Delta T_{g}}{T_{g}(\Delta\log\omega)}$ and is estimated to be 0.025. In comparison, the value of $k$ parameter in Ni rich NiTi alloys is about 0.02 \cite{Sarkar200595}. This slightly higher value of $k$ parameter in Ni$_2$Mn$_{1.4}$Fe$_{0.1}$In$_{0.5}$ could be due to presence of larger sized domains in the present alloy as compared to those in Ni rich NiTi alloys. Another interesting aspect to be noted is the presence of a smaller but distinct feature in temperature dependence of $\tan\delta$. This feature appears between 375K to 400K (marked by arrow in Fig. \ref{fig:VF}) and matches with the weak first order transition seen in the resistivity measurements on this alloy.


\begin{figure}[h]
\begin{center}
\includegraphics[width=\columnwidth]{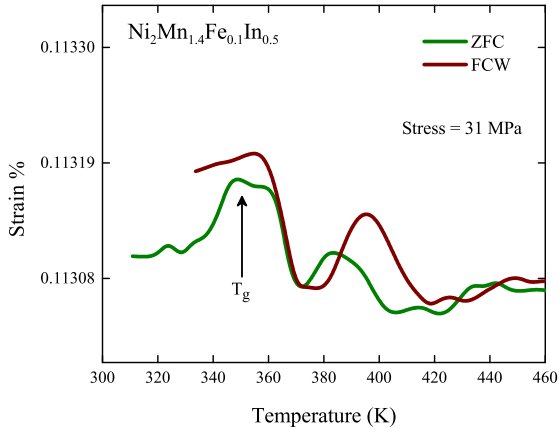}
\caption{\% Strain as a function of temperature recorded during zero field cooled and field cooled cycles at 6Hz.}
\label{fig:FCZFC}
\end{center}
\end{figure}

To further check the presence of strain glassy phase in Ni$_2$Mn$_{1.4}$Fe$_{0.1}$In$_{0.5}$, history dependence of strain during zero field cooled (ZFC) and field cooled (FC) cycles was carried out and the results are presented in Fig. \ref{fig:FCZFC}. A clear deviation between the two curves which is considered as a critical proof of existence of strain glassy phase, can be seen from  $\sim$ 363 K which is above the $T_g$ = 350K. This confirms that Ni$_2$Mn$_{1.4}$Fe$_{0.1}$In$_{0.5}$ is indeed a strain glass.

The question then arises about the room temperature structure of Ni$_2$Mn$_{1.4}$Fe$_{0.1}$In$_{0.5}$ and the presence of a first order transition in its resistivity measurement. For a strain glass the structure is expected to be invariant across the transition. However, the $x = 0.1$ alloy exhibits a martensitic structure. In magnetic cluster glasses, there are examples of materials exhibiting glassy characteristics and yet presenting long range magnetic order. Recently such a phenomenon has been explained to be due to presence of clusters large enough to show characteristics of long range magnetic order in neutron diffraction but still exhibit glassy behavior due to limited interaction between clusters \cite{DiasPRB2017}. Presence of such large strain domains separated by non martensitic regions in Ni$_2$Mn$_{1.4}$Fe$_{0.1}$In$_{0.5}$ cannot be ruled out. Structural studies on higher Fe content ($x \ge 0.15$) alloys show presence of two, austenitic and martensitic, structural phases. Such a scenario, wherein either the concentration of the cubic austenitic phase is quite low or the grains are not large enough to be detected in XRD, could be also present in $x$ = 0.1 composition. It is in fact supported by the presence of a weak but distinct features corresponding to a first order transition in resistivity and ac storage modulus and loss measurements. It appears that Ni$_2$Mn$_{1.4}$Fe$_{0.1}$In$_{0.5}$ consists of clusters that are largely deficient in Fe and hence undergo martensitic transition at a temperature very close to that of undoped alloy. These clusters are large enough to show signatures of incommensurate modulated 7M structure in XRD but have very limited interactions with other similar clusters due to presence of minor Fe rich impurity phases. The slightly higher value of $k$ parameter (0.025) as compared to that observed in NiTi alloys also supports the presence of large clusters. Such an unusual strain glassy phase reported here needs to be investigated further using temperature dependent structural and local structural techniques.

In conclusion, Fe doping in Ni$_2$Mn$_{1.5-x}$Fe$_x$In$_{0.5}$ results in reduction of martensitic transition temperature with increasing $x$ as evidenced from exothermic and endothermic peaks during warming and cooling cycles in DSC. Though at room temperature Ni$_2$Mn$_{1.4}$Fe$_{0.1}$In$_{0.5}$ exhibits incommensurate, modulated 7M martensitic structure, its DSC thermograms show a nearly vanishing feature indicating presence of short range ordering of strain vector. A frequency dependent behavior of storage modulus and loss that obeys Vogel-Fulcher law and the presence of ergodicity breaking between zero field cooled and field cooled strain measurements just above the transition temperature confirm presence of a frozen glassy state below $T_g$ = 350K. Despite martensitic structure, presence of a strain glassy phase can be explained to be due to presence of strain domains which are large enough to present signatures of long range martensitic order in diffraction but remain non interacting with each other due to presence of impurity phases rich in Fe.

\section*{Acknowledgements}
The authors wish to acknowledge the financial assistance from Science and Engineering Research Board, Govt. of India under the project SB/S2/CMP-0096/2013.

\bibliographystyle{apsrev4-1}
\bibliography{Ref}

\end{document}